# Temperature-insensitive fused-tapered fiber couplers based on negative thermal expansion material coating

ZE-LONG HUANG,[a] JIE XU,[a,*] JUE LI,[a] CHUN-ZHAO MA,[a] JIAN LUO,[a] XIN YU,[a] YUN-QIAO HU,[a] CHANG-LEI GUO,[a] AND HSIEN-CHI YEH[a]

[a]MOE Key Laboratory of TianQin Mission, TianQin Research Center for Gravitational Physics & School of Physics and Astronomy, Frontiers Science Center for TianQin, CNSA Research Center for Gravitational Waves, Sun Yat-sen University (Zhuhai Campus), Zhuhai 519082, China

A B S T R A C T

A new method based on negative thermal expansion material coating is proposed to realize temperature-insensitive fiber coupler. By coating a layer of modified epoxy resin with a negative thermal expansion coefficient onto the coupling region of fiber coupler, a stable splitting ratio over a wide temperature range can be achieved. A finite-element model for simulating the influence of thermal fluctuations on fused-tapered fiber coupler's splitting ratio is built and verified via experimental test. Furthermore, using this model, the influence of the thickness, length, and thermal expansion coefficient of the coating material on the splitting ratio is studied. Through adjusting the parameters of the coating, the temperature stability of the fiber coupler splitting ratio can be improved by more than one order of magnitude and improved to $1.2 \times 10^{-5}$ /K. The temperature-insensitive fused-tapered fiber coupler can find important application in optical precision measurement under extreme temperature environment, such as inter-satellite laser interferometry and high-precision fiber gyroscopes.

## 1. Introduction

Ultra-stable optical systems are of great importance in gravitational wave detection [1-4], precision optical measurements [5], and cold atomic physics [6, 7]. In ultra-stable low noise laser systems, beam splitters are key elements for outputting and monitoring power, and the change of the splitting ratio will directly affect the laser's intensity noise at low frequency band [7, 8]. Even if active power stabilization units are applied, since the output port is always outside the power feedback loop, the fluctuations of splitting ratio will still affect the power noise [9]. Therefore, high-stability beam splitters are in urgent need. Currently, typical optical beam splitters mainly include free-space optical beam splitters [7] and fiber couplers [10]. The splitting ratios of these components are easily affected by temperature due to thermal expansion and thermo-optic effects. To improve the stability of the splitting ratio, the most used method is adding active temperature control for the beam splitter [9], which will increase the complexity and cost, as well as the failure rate of the device.

The concept of using two or several materials with different optical properties to achieve an expected optical property has been widely used in thin film optics [11-13], optical dispersion control [14, 15], and optical sensors [16-18]. For example, optical micro-resonators with high flat dispersion have been achieved through the combination of several materials, thereby realizing broadband nonlinear optical effects (such as Kerr optical frequency combs [19-22]). Using composite materials with unique optical properties, the sensitivity of optical sensors can be adjusted to enhance sensitivity to specific characteristics [23-25]. A fused-tapered fiber coupler typically uses standard single-mode fibers, which are mainly made of fused silica with low refractive index for the cladding and high refractive index for the core. An optical clear adhesive is used to protect the taper zone. Under normal laboratory temperature (20 ~ 25 °C), the optical splitting ratio fluctuation is as high as $10^{-4}$ /K, which is far from dedicated design for ultra-stable laser applications requiring high power stabilities [26-28].

In this work, we propose a fused-tapered fiber coupler (FTFC) with extremely low temperature sensitivity via adding a layer of negative thermal expansion material. We first introduce the basic working principle of the FTFC. Both the theoretical and finite-element models are built for splitting ratio simulations, which are verified via experimental tests. We then introduce a material with negative thermal expansion to finely optimize the sensitivity of splitting ratio with temperature, where parameters of thickness, thermal expansion coefficient and length of the coating material are investigated. The simulation results show that the sensitivity of splitting ratio with temperature can reach as low as $1.2 \times 10^{-5}$ /K by using a layer of modified epoxy resin material, which has been improved by more than one order of magnitude. This FTFC with extremely low temperature sensitivity will significantly reduce the power noise added to an ultra-stable laser, thereby enhancing the performance of laser optics for gravitational wave detection in space and other precision measurement activities [29-33].

* Corresponding author.
E-mail address: xujie63@mail.sysu.edu.cn

## 2. Fused-tapered fiber couplers

The FTFC is a coupling system consisting of two optical fibers that are tapered, as shown in Fig. 1. In a single-mode fiber waveguide [34], the fiber core can confine most of the light, but due to the existence of the evanescent field, a small amount of light leaks into the cladding. As the optical fiber is tapered, the cores become thinner and closer, the normalized frequency $V$ decreases [35], and the constraint of the core on the transmission of the fundamental mode also decreases. Finally, the portion of optical power leaked to the cladding increases. Since the claddings of the two fibers in the coupling region merge, the two cores can exchange energy through the evanescent field via the merged cladding area, eventually completing coupling between the two fibers.

The splitting ratio of the FTFC can be adjusted by controlling the fiber length in the coupling region during fusion [36], and its theoretical expression is:

$$CR = P_1 / (P_1 + P_2) = \cos^2 Kz \tag{1}$$

$$K = \frac{1}{k_0} \cdot \frac{U^2}{a^2 V^2} \cdot \frac{K_0(Wd/a)}{K_1^2(W)} \tag{2}$$

$$U = (n_1 k_0^2 - \beta^2)^{1/2} \cdot a \tag{3}$$

$$W = (\beta^2 - n_2 k_0^2)^{1/2} \cdot a \tag{4}$$

Where $CR$ is the splitting ratio of the fiber coupler. $P_1$ and $P_2$ represents the output power of port 1 and port 2, respectively. $z$ is the length of the coupling region. $K$ is the coupling coefficient of the two cores in the coupling region. $V = 2\pi\sqrt{n_1^2 - n_2^2}$, and $a$ is the core radius of the fiber in the coupling region. $\lambda_0$ is the wavelength of the light wave. $n_1$ and $n_2$ are the refractive index of the core and the refractive index of the cladding, respectively. $d$ is the distance between the two cores in the coupling region. $U$ is the normalized radial phase constant of the field. $W$ is the normalized attenuation constant of the field, and $\beta$ is the propagation constant of the fiber. $K_0$ and $K_1$ are the zero-order and first-order modified second-order Bessel functions, respectively, with $k_0 = 2\pi/\lambda_0$.

According to the above formula, the main factors affecting the splitting ratio are the coupling coefficient $K$ and the length of the coupling region $z$. The coupling coefficient is related to the propagation constant $\beta$ and the fiber core radius of the coupling region $a$. However, the propagation constant $\beta$ depends on the normalized frequency of the fiber [37]. Therefore, the main factors affecting the splitting ratio of the fiber coupler can be summarized as $z$, $a$, $d$, $n_1$ and $n_2$. The theoretical expression of the change of splitting ratio with temperature is:

$$\frac{\partial CR}{\partial T} = -2\cos(K[a,n_1,n_2,d]z)\sin(K[a,n_1,n_2,d]z)$$
$$K[a,n_1,n_2,d]\frac{\partial z}{\partial T} + z(K[n_1]\frac{\partial n_1}{\partial T} + K[n_2]\frac{\partial n_2}{\partial T} + \tag{5}$$
$$K[a]\frac{\partial a}{\partial T} + K[d]\frac{\partial d}{\partial T})$$

After a single-mode fiber is fused and tapered, the expression of the optical fiber core radius in the coupling region [36] is:

$$a(z) = a_0 \exp((\frac{L_t}{2\pi \cdot L_s}) \cdot [\sin(2\pi z / L_t - \pi/2)]) \tag{6}$$

In the formula, $a_0$ is the fiber radius before fusing, $L_t$ is the total length of the fused-taper region, and $L_s$ is the length of the dynamic fused-tapered region. In the following section, we will apply equation (6) to build a finite-element simulation model of FTFC.

## 3. Simulation and experimental test of temperature fluctuation introduced splitting ratio change

We used finite-element analysis software to simulate the fiber coupler with a two-dimensional model. Firstly, we established the geometric model of the FTFC. The geometry shape of the fiber in the coupling region after fused taper is given by equation (5). Taking commercial 50:50 fiber coupler (AFR, PMC-1-1064-50-P-2) as an example, we defined the bending radius of the fiber in the transition region (see Fig. 1) as 2.5 mm. The parameters and values involved in the model are shown in table 1.

**Table 1.**

The parameters and values involved in the model

| Parameters | Describe | Value |
| --- | --- | --- |
| $a_0$ | Core diameter | 5.5 μm |
| $b$ | Cladding diameter | 125 μm |
| $n_1$ | Core refractive index | 1.4551 |
| $n_2$ | Cladding refractive index | 1.4500 |
| $\gamma_1$ | Core thermo-optical coefficient | $6.8\times10^{-6}$ /K |
| $\gamma_2$ | Cladding thermo-optical coefficient | $6.8\times10^{-6}$ /K |
| $\lambda$ | Wavelength | 1064 nm |
| $L_x$ | Length of coupling region | 26 mm |
| $d$ | Core separation distance | 7.8 μm |
| $\alpha$ | Thermal expansion coefficient | $5.5\times10^{-7}$ /K |

In the model, we input an optical signal into one of the waveguide arms, as shown in Fig. 1(a). When the optical signal arrives at the coupling region, it will be coupled between two fiber core waveguides. We calculate the time-averaged Poynting vector S of the output port 1 and port 2 to obtain the optical power variation of the two ports in the fiber coupler, and substitute it into equation (1) to obtain the corresponding splitting ratio.

Using this model, we calculated the variation of the splitting ratio with coupling length at room temperature of 20 °C, as shown in Fig. 1(c). The splitting ratio at the output of the fiber coupler changes sinusoidally with the length of coupling region, which is consistent with the theoretical prediction of equation (1). Through the model, we can also clearly see that the propagated light waves in the coupling region are well-confined in the cladding and passes between the cores, as shown in Fig. 1(b).

In the next stage, we will use this model to further investigate the effect of temperature fluctuation on the splitting ratio of the FTFC.

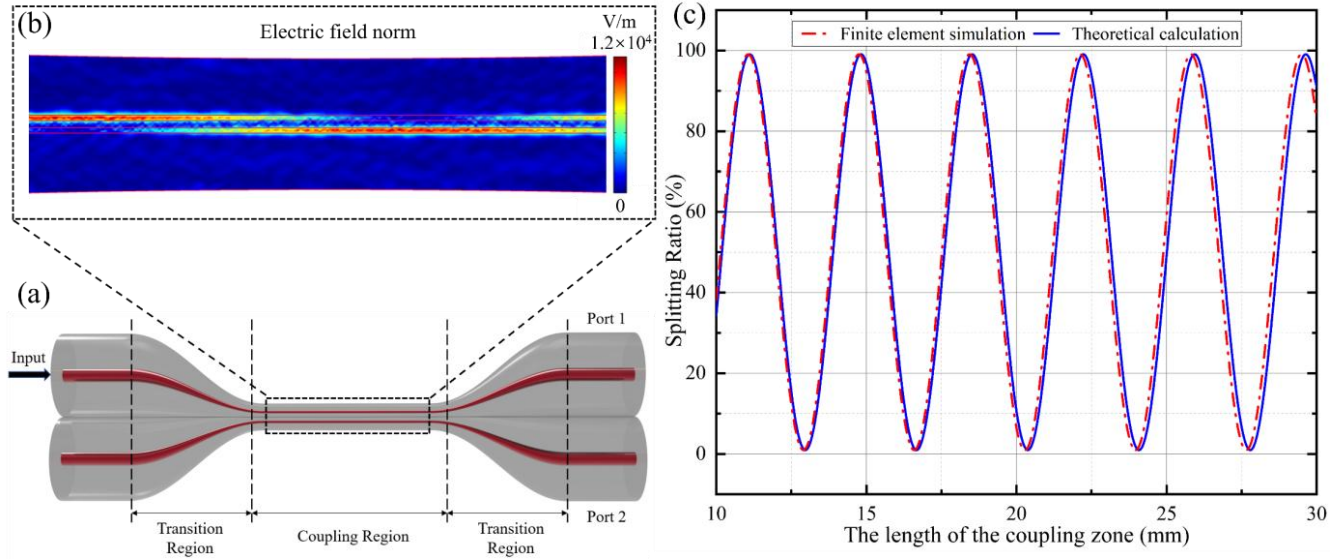

**Fig.1** FTFC model; (a) the geometry of the FTFC; (b) colormap of electric field in the coupling region from finite-element simulation; (c) splitting ratio versus length of coupling zone. Blue solid line: analytical calculation using formulas (1), red dashed line: numerical calculation using finite-element simulation.

As mentioned in section 2, the main factors affecting the splitting ratio of the fiber coupler are the length of the coupling region $z$, the core radius of the coupling region $a$, the distance between the two cores in the coupling region $d$, the core refractive index $n_1$, and the cladding refractive index $n_2$. These parameters are affected by the ambient temperature. When the ambient temperature changes, due to thermal expansion and contraction, $z$, $a$ and $d$ will change [38]. On the other hand, due to the thermo-optical effect, $n_1$ and $n_2$ change [39]. In the model, we take the thermal expansion effect and the thermo-optical effect into account. By changing the ambient temperature, we obtain the curve of the fiber coupler splitting ratio changing with temperature. Finally, to verify the accuracy of the model, the simulated results are compared with the experimental results.

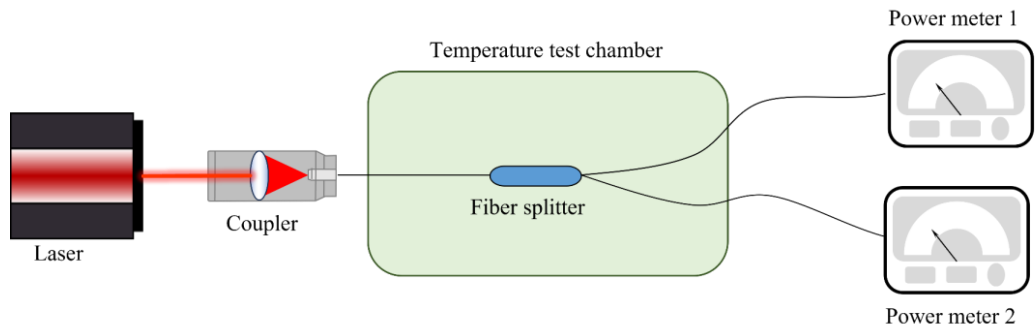

**Fig.2** Experimental schematic for testing the splitting ratio versus temperature.

As shown in Fig. 2, we used a single-frequency laser with high power-stability at 1064 nm as the light source. The fiber coupler (AFR, PMC-1-1064-50-P-2) was placed in a temperature test chamber (DHT, DHT260-V2) with an adjustable temperature of 0 ~ 60 °C. The optical output power of the fiber coupler is measured by two power meters (Thorlabs, PM400). The splitting ratio of the fiber coupler was 46.3% at room temperature. The experimental results are shown in Fig. 3. The results show that the splitting ratio of the fiber coupler is sensitive to temperature fluctuations. In the temperature range of 0 ~ 45 °C, the splitting ratio of the fiber coupler decreases as the temperature increases, and changes nearly linearly in the 0 ~ 30 °C range. When the temperature rises above 45 °C, the trend of change reverses, and the splitting ratio shows an increasing trend as the temperature rises. The finite element simulation results were in good agreement with the experimentally measured results.

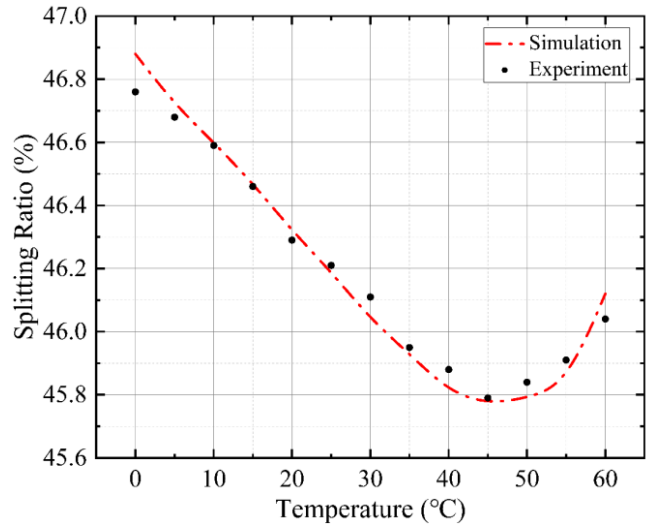

**Fig. 3 Splitting ratio versus temperature, simulated and tested with a commercial fiber coupler. Red dashed line is the finite-element simulation results. Black dots are from experimental measurements.**

In general, by comparing the finite element simulation results of the fiber coupler with the theoretical prediction and experimental test results, we confirmed the reliability of our FTFC model. Moreover, both the simulation and the experimental results show that the splitting ratio of the fiber coupler is highly sensitive to temperature fluctuations, which will greatly affect the power stability of the fiber coupler's output signal.

## 4. Design of fused-tapered fiber couplers with high temperature stability

In order to reduce the temperature sensitivity of the fiber coupler and achieve high temperature stability output, we proposed a method to improve the stability of the fiber coupler splitting ratio by coating a layer of negative thermal expansion material onto the coupling region. The method was verified via the finite-element model we established. Based on the above fiber coupler model, a layer of modified epoxy resin adhesive with negative thermal expansion coefficient is coated on the coupling region to compensate for the thermal expansion and thermo-optical effect caused by the change of ambient temperature, so as to achieve a highly stable splitting ratio. The schematic of model is shown in Fig. 4.

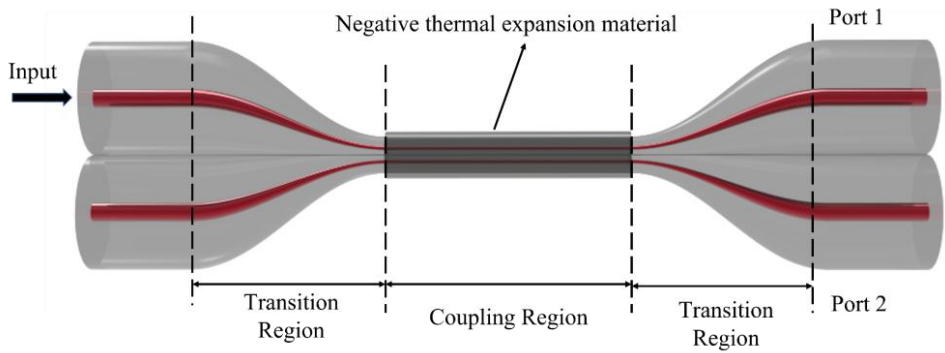

**Fig. 4 Schematic diagram of high temperature stability fiber coupler with negative thermal expansion material coating**

We assume that the applied modified epoxy resin adhesive layer is uniform and covers the entire coupling region, with a thermal expansion coefficient is $-7.43\times10^{-5}$ /K [40]. We chose room temperature of (20 °C) ± 10 °C as the working range. Firstly, we studied how the thickness of the coating layer affects the splitting ratio under different temperature via the same simulation model (discussed in Section 3) with an extra coating in the coupling region. The simulation results are shown in Fig. 5(a). The results indicate that the change of the splitting ratio of the fiber coupler with temperature can be greatly reduced by selecting an appropriate thickness of the modified epoxy resin adhesive layer. When the coating thickness is 28 μm, the temperature sensitivity of the fiber coupler splitting ratio is reduced from $3\times10^{-4}$ /K to $1.2\times10^{-5}$ /K, and the temperature stability of the fiber coupler splitting ratio is improved by more than one order of magnitude.

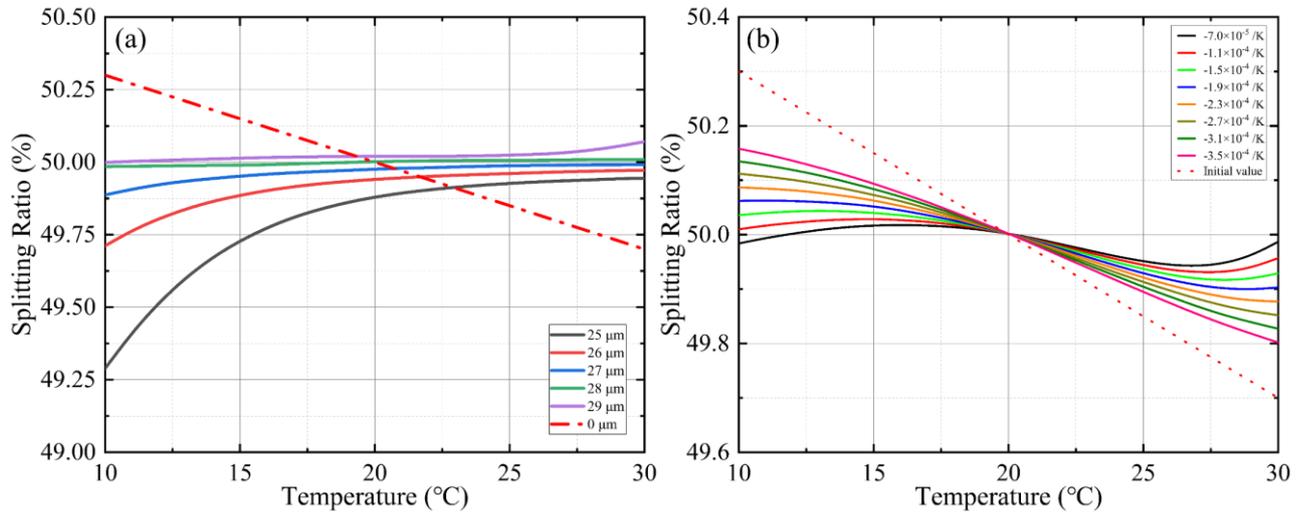

**Fig. 5 Finite-element simulation results of the FTFC with negative thermal expansion material coating: (a) Splitting ratio versus temperature with change of coating thickness; (b) Splitting ratio versus temperature with change of thermal expansion coefficient of coating material. The other parameters are set at their nominal values during the simulations.**

Then, we studied how the thermal expansion coefficient of the coating layer affects the splitting ratio under different temperatures. Fig. 5(b) shows the simulated change in splitting ratio with temperature after the coupling region of fiber coupler is coated with modified epoxy resin with different thermal expansion coefficients, while the coating thickness is set to 28 μm. In the simulation, epoxy resin with negative thermal expansion coefficient ranging from $-7\times10^{-5}$ /K to $-3.5\times10^{-4}$ /K was used. According to the simulation results, as the negative thermal expansion coefficient of the epoxy resin material increases, the temperature sensitivity of the fiber coupler splitting ratio also increases. However, in the range of 10 to 30 °C, the temperature stability of the splitting ratio of the fiber coupler coated with modified epoxy resin is better than that of the uncoated one.

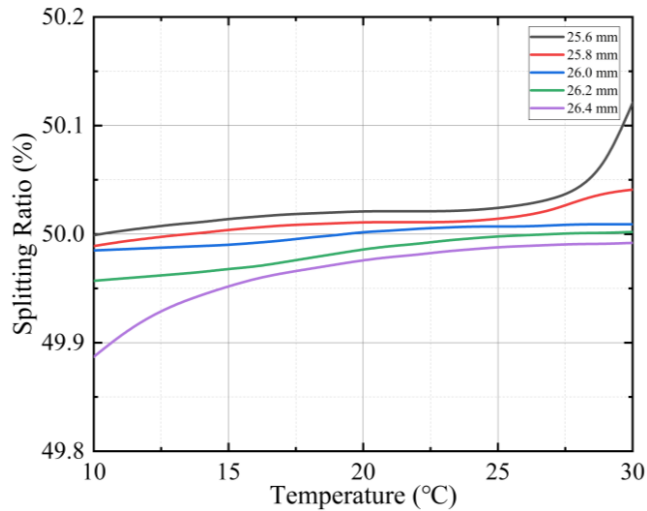

**Fig. 6 Effect of modified epoxy resin coating length on splitting ratio of fiber coupler with temperature**

In addition, we also studied how the coating length of the modified epoxy resin affects the splitting ratio of fiber coupler under different temperatures. We set the coating thickness to 28 μm and the thermal expansion coefficient of the modified epoxy resin to $-7.43\times10^{-5}$ /K. As shown in Fig. 6, when the coating length changes within ± 0.2 mm of the coupling zone length, the temperature stability of the fiber coupler splitting ratio also changes greatly. Moreover, the greater the coating length deviates from the coupling zone length, the worse the temperature stability becomes. Therefore, in practice, we need to control the coating length accuracy of the negative thermal expansion material better than ± 0.2 mm.

Finally, we would like to briefly discuss the feasibility of coating negative thermal expansion materials onto the coupling region of fiber couplers. The negative thermal expansion material mentioned in this paper has good tensile strength, and its elasticity modulus is close to that of adhesive commonly used for fiber coating, which can be fully cured after 12 hours at 60°C [40]. Therefore, the negative thermal expansion material could be coated onto the coupling zone of the FTFC. Moreover, the thermal expansion coefficient of the negative thermal expansion material can be adjusted by controlling the doping ratio of polyamide [40]. In terms of the coating process, we could put the coupling zone of the prepared FTFC into a customized fine mold, and pour the modified epoxy adhesive into the mold, and the adhesive can be cured to obtain temperature-insensitive FTFC. The process is very similar to the current fiber coating process, where the length and thickness of the coating can be controlled by customizing coating molds.

## 5. Conclusion

In this paper, we introduced a method to reduce the temperature sensitivity of fused-tapered fiber coupler's splitting ratio by coating a layer of modified epoxy resin with a negative thermal expansion coefficient onto the coupling region of the fiber coupler. Through finite-element modeling, we systematically study the effect of the coating material's thickness, length, and thermal expansion coefficient on the fiber coupler's temperature sensitivity. Moreover, the simulation results showed that for a 50:50 fiber coupler with a coupling region length of 26 mm, after coating a layer of modified epoxy resin with a length of 26 mm, a thickness of 28 μm, and a thermal expansion coefficient of $-7.43\times10^{-5}$ /K, the temperature stability of the fiber coupler splitting ratio can reach $1.2\times10^{-5}$ /K, which represents an improvement of more than one order of magnitude higher than the uncoated case. These fiber couplers with high temperature stability are of great significance in the detection of gravitational waves and the application of high-precision fiber gyroscopes.

## Funding

This work was supported by National Key Research and Development Program of China (2023YFC2205504), National Natural Science Foundation of China (12404489), Fundamental Research Funds for the Central Universities, Sun Yat-sen University (24QNPY162), and Major Projects of Basic and Applied Basic Research in Guangdong Province (2019B030302001).

## CRediT authorship contribution statement

**Zelong Huang**: Writing-original Draft, Writing-review & editing, Methodology, Formal analysis, Data curation. **Jie Xu**: Writing-review & editing, Formal analysis, Methodology, Supervision, Resources, Project administration, Funding acquisition. **Jue Li**: Formal analysis, Writing-review & editing. **Chunzhao Ma**: Formal analysis, Writing-review & editing. **Jian Luo**: Formal analysis, Writing-review & editing. **Xin Yu**: Formal analysis, Writing-review & editing. **Yunqiao Hu**: Formal analysis, Writing-review & editing. **Changlei Guo**: Investigation, Methodology, Supervision, Resources, Formal analysis, Writing-review & editing, Funding acquisition. **Hsien-Chi Yeh**: Supervision, Resources, Investigation, Funding acquisition.

## Declaration of competing interest

The authors declare that they have no known competing financial interests or personal relationships that could have appeared to influence the work reported in this paper.

## Data availability

Data will be made available on request.